\documentstyle [12pt] {article}

\catcode`\@11
\def\section{
    \setcounter{equation}{0}
\@startsection {section}{1}{\z@}{-3.5ex plus -1ex minus -.2ex}
{2.3ex plus .2ex}{\large\bf}
}
\catcode`\@13

\newcommand{\nummer}[1]{\hskip 12 true cm #1 \par}
\newcommand{\netnum}[1]{\vspace{-14pt}\hskip 12 true cm #1 \par}
\newcommand{\monat}[1]{\hskip 12 true cm #1
                       \par \vspace*{1 cm}}
\newcommand{\titel}[1]{{\renewcommand{\thefootnote}{\fnsymbol{footnote}}
                       \Large\bf\vskip 0 true cm
                       \begin{center}#1\end{center}
                       \setcounter{footnote}{0}}
                       \normalsize\vskip 1.2 true cm}
\newcommand{\autor}[1]{{
                       \renewcommand{\thefootnote}{\arabic{footnote}}
                       \begin{center} {\large #1 }\end{center}}
                       \setcounter{footnote}{0}}
\newcommand{\adresse}[1]{\vspace*{-1.1 true cm}\begin{center} {\it #1 }
                         \end{center}
                         \vskip 0.5cm}

\newcommand{\bye}{\end{document}}
\newcommand{\be}{\begin{equation}}
\newcommand{\ee}{\end{equation}}
\newcommand{\bes}{\begin{eqnarray}}
\newcommand{\ees}{\end{eqnarray}}
\newcommand{\ema}{\end {array} \right)}
\newcommand{\pslash}{\kern 0.1 em p\kern -0.45em /}
\newcommand{\dslash}{\kern 0.1 em \partial\kern -0.55em /}
\newcommand{\sla}[1]{\kern 0.1 em #1\kern -0.55em /}
\newcommand{\ra}{\rightarrow}

\newcommand{\lra}{\longrightarrow}

\newcommand{\R}{\mbox{I\kern -0.22em R\kern 0.30em}}
\newcommand{\N}{\mbox{I\kern -0.22em N\kern 0.30em}}
\newcommand{\C}{\mbox{\kern 0.20em \raisebox{0.09ex}{\rule{0.08ex}{1.22ex}}
                \kern -0.60em C\kern 0.30em}}
\newcommand{\Z}{\mbox{\sf Z\kern -0.40em Z\kern 0.30em}}

\newcommand{\OD}{\mbox{$\Omega_D$} }
\newcommand{\cA}{\mbox{$\cal A$} }
\newcommand{\cH}{\mbox{$\cal H$} }
\newcommand{\cJ}{\mbox{$\cal J$} }
\newcommand{\cM}{\mbox{${\cal M}$} }

\newcommand{\cL}{\mbox{${\cal L}$} }
\newcommand{\cD}{\mbox{${\cal D}$} }
\newcommand{\cE}{\mbox{${\cal E}$} }
\newcommand{\bo}[1]{\mbox{\boldmath$#1$}}
\newcounter{zaehler}
\newenvironment{iliste}{\begin{list}{(\roman{zaehler})\hfill}{
 \usecounter{zaehler}\topsep-3pt\partopsep0pt\labelwidth0.8cm\leftmargin1.3cm
 \labelsep0.2cm\rightmargin0cm\parsep0.5ex plus0.2ex minus0.1ex
 \itemsep0ex plus0.2ex}}{\end{list}}
\begin{document}
\begin{titlepage}
\nummer{MZ-TH/93-38}
\netnum{gr-qc/9312031}
\monat{December 1993}
\titel{Gravity, Non-Commutative Geometry and the Wodzicki Residue\footnote{
{\rm Work supported in part by the PROCOPE agreement between the
University of Aix-Marseille and the Johannes Gutenberg-Universit\"at of Mainz.
}}}
\autor{W.~Kalau\footnote{e-mail: kalau{\char'100}vipmza.physik.uni-mainz.de},
M.~Walze\footnote{e-mail: walze{\char'100}vipmza.physik.uni-mainz.de}}
\adresse{Johannes Gutenberg Universit\"at\\
Institut f\"ur Physik\\
55099 Mainz}
\begin{abstract}
We derive an action for gravity in the framework of non-commutative
geometry by using the Wodzicki residue. We prove that for a Dirac operator
$D$ on an $n$ dimensional compact Riemannian manifold with $n\geq 4$,
$n$ even, the Wodzicki residue Res$(D^{-n+2})$ is the integral of the second
coefficient of the heat kernel expansion of $D^{2}$. We use this result to
derive a gravity action for commutative geometry which is the usual Einstein
Hilbert action and we also apply our results to a non-commutative extension
which, is given by the tensor product of the algebra of smooth functions on a
manifold and a finite dimensional matrix algebra. In this case we obtain
gravity with a cosmological constant.
\end{abstract}
\end{titlepage}
\section{Introduction}
Although General Relativity is well established as a classical theory
of gravitational interaction we still do not know how to describe gravity
at distances of order Planck-length, i.e. we do not have a theory of
gravity which is compatible with the quantum theory of the other fundamental
interactions which are experimentally well understood in the framework of
the Standard Model.

Since a considerable amount of effort has been spent on this problem one
may draw the conclusion that the mathematical concepts of General Relativity
have to be changed, or more precisely, our classical geometrical concepts
may not be well suited for the description of gravity at small distances.
A promising direction seems to be the generalization of
geometry to non-commutative geometry \cite{cobuch} which has been used
by A.~Connes and J.~Lott \cite{colo} to derive a model for the electroweak
interactions. This led to a new interpretation
of the Higgs particle as a connection on a discrete space. The geometrical
set-up studied in \cite{colo} is a tensor product of a compact Riemannian
manifold and a discrete two-point space. This geometry leads to theories
with one symmetry breaking scale. A generalization to geometries where the
discrete space has more than two points was performed in \cite{chams1} in
order to describe theories with several symmetry breaking scales like
GUT-models. There is also an alternative approach to non-commutative
geometry, which allows for the same geometrical picture, i.e.
smooth manifold $\times$ discrete space, but it is not restricted to
compact Riemannean space-time. This approach was developed in
\cite{robert,florian,hps2}. A comparison of both models can be found in
\cite{PPS}.

Now it seems to be only natural
to apply the concepts of non-commutative geometry to gravity since it is a
theory of space-time geometry. A reformulation of gravity in this language
brings us in the position to use the power of non-commutative geometry which
might be of some use to find a consistent quantum theory for gravity.

A first step in this direction was made by Chamseddine et al.~\cite{chams} who
generalized the notion of cotangent space to the case where the geometry
is given by a tensor product of a smooth manifold and a discrete two-point
space. They studied a vielbein and a
connection which are related by generalized Cartan structure equations. This
led to gravity coupled to a scalar field.

However, in this article we will follow a different line of approach. We shall
use the fact that the choice of a K-cycle over an algebra specifies the metric
properties of the `manifold' described by the algebra. More precisely, the
geometric structure is encoded in the Dirac operator of a K-cycle and
therefore we will use this fact to derive an action.

We start in the next section with a brief introduction to the general concepts
of non-commutative geometry. The Dixmier trace as an operator theoretic
substitute is introduced in sec.3 and we discuss its relation to the heat
kernel
expansion. We use this as a motivation to derive a gravity action by
selecting the second coefficient of the heat kernel expansion. The proof of
our main result, namely that the Wodzicki residue of $D^{-n+2}$, where $D$
is a Dirac operator on an n dimensional compact Riemannian manifold (n even),
picks out the second heat kernel coefficient is presented in sec.4. In sec.5
we apply our result to the usual Dirac operator without and with torsion.
In this case we obtain the Einstein Hilbert action. We also consider a simple
extension to non-commutative geometry
which is given by the tensor product of the algebra of smooth functions
on a manifold and a finite dimensional matrix algebra. Such algebras are
used in model building \cite{colo,chams1} and also in \cite{chams}.
For those geometrical set-ups we obtain a gravity action with a
cosmological constant and therefore our result is different from that
obtained in \cite{chams}. We end this article with some conclusions in sec.6.
\section{Dirac-K-Cycles and Metric Structures}
In this section we briefly review some of the main concepts of non-commutative
geometry in order to make this article self contained and to fix our
notation. For a more comprehensive presentation of this subject
we refer to \cite{cobuch,VaBo}.

Let \cA be an associative unital algebra. We can construct a bigger algebra
$\Omega\cA$ out of it by associating to each element $a\in\cA$ a symbol
$\delta a$. $\Omega\cA$ is the free algebra generated by the symbols $a$,
$\delta a$, $a\in\cA$ modulo the relation
\be
\delta (ab)=\delta a\, b + a\delta b\;\; .\label{univrel}
\ee
With the definition
\bes
\delta(a_0\delta a_1\cdots\delta a_k)&:=&\delta a_0\,\delta a_1\cdots\delta a_k
\;\; ,\\
\delta(\delta a_1\cdots\delta a_k)&:=& 0
\ees
$\Omega\cA$ becomes a $\Z\!$-graded differential algebra with the odd
differential $\delta$ and $\delta^2=0$. $\Omega\cA$ is called the universal
differential envelope of \cA.

The next element in this formalism  is a K-cycle $(\cH, D)$ over \cA, where
\cH is a Hilbert space such that there is an algebra homomorphism
\be
\pi : \cA \lra B(\cH)\;\; ,
\ee
where $B(\cH)$ denotes the algebra of bounded operators acting on \cH. $D$ is
an unbounded self-adjoint operator with compact resolvent
such that $[D,\pi(a)]$ is bounded for all $a\in \cA$. It
is the triple $(\cA ,\cH ,D)$ which contains all geometric information.

We can use $D$ to extend $\pi$ to an algebra homomorphism of $\Omega\cA$
by defining
\be
\pi(a_0\delta a_1\cdots\delta a_k):=\pi(a_0)[D,\pi(a_1)]\cdots[D,\pi(a_k)]
\;\; . \label{Drep}
\ee
However, in general $\pi(\Omega\cA)$ fails to be a differential algebra.
In order to repair this, one has to divide $\Omega\cA$ by the two sided
$\Z\!$-graded differential ideal \cJ given by
\be
\cJ = \bigoplus_{k\in\N} \cJ^k\;\;\; ,\;\;\; \cJ^k:=(\hbox{ker}\pi)^k +
\delta(\hbox{ker}\pi)^{k-1}\;\;\; .
\ee
Now we can define the non-commutative generalization of the de Rham algebra,
$\Omega_D\cA$, as
\be
\OD\cA := \bigoplus_{k\in\N} \pi(\Omega^k\cA) /\pi(\cJ^k)
\ee
$\OD\cA$ is an $\Z\!$-graded differential algebra, where the differential $d$
is defined by
\be
d[\pi(\omega)] := [\pi(\delta\omega)]\;\;, \;\; \omega\in\Omega\cA\;\;.
\ee
If we take, for example, $\cA = C^\infty(M)$,
the algebra of smooth functions on a compact Riemannian
spin manifold $M$, \cH as the space of square-integrable spin-sections and
the Dirac operator $D=\dslash$ then $\OD\cA$ is the usual de Rham
algebra \cite{cobuch}.

A remarkable fact is that the geodesic distance $d(p,q)$ on such a manifold $M$
for any $p,q \in M$ is encoded in the Dirac operator $D$
(the algebra is $C(M)$):
\be
d(p,q)= \mbox{sup} \{|a(p)-a(q)| ;\; a\in C(M),\; ||\, [D,a]\, ||\leq 1\}
\label{geodis}
\ee
No arcs are involved on the right hand side of this relation and therefore
eq.(\ref{geodis}) can be taken as a definition of geodesic distance which
still makes sense in situations where arcs cannot be defined.
We use this as a motivation to construct an action which only depends on
the choice of the Dirac operator for a K-cycle.
\section{Dixmier Trace and Heat Kernel Expansion}
In order to write down an action in the operator theoretic language we
need a functional which replaces integration. For Yang-Mills theory
the correct substitution is given by the Dixmier trace \cite{cobuch}. It
is the unique extension of the ususal trace to the class
${\cal L}^{(1,\infty)}({\cal H})$ \cite{dix} which is an ideal in the algebra
of bounded operators. The elements of this ideal are characterized by the
condition that for any $T\in{\cal L}^{(1,\infty)}({\cal H})$ the ordered
eigenvalues $\lambda_i$
of $|T|$ satisfy
\be
\sup_{N}{1\over\log N}\sum_{i=0}^N \lambda_i < \infty\;\; .
\ee
On this ideal the Dixmier trace is defined as
\be
Tr_\omega(T) = \lim_{N\ra\infty}{1\over\log N}\sum_{i=0}^{N-1} \lambda_i\;\; .
\ee
An important result in non-commutative geometry is the trace theorem
of A.~Connes \cite{co} which states that
\be
Tr_\omega (T) = \lim_{p\ra 1+}(p-1)\zeta_T(p)\; , \;\;
T\in {\cal L}^{(1,\infty)}({\cal H})\;\;, \label{trateo}
\ee
with
\be
\zeta_T(p) = tr(T^p)\;\;.
\ee
If we now take a K-cycle $({\cal A},{\cal H}, D)$ where $\cal H$ denotes the
space of square-integrable sections of a Clifford
module\footnote{In the algebraic language $\cal H$ is a finitely generated
projective module over $\cal A$.} on a compact Riemannian manifold $M$ of
dimension $n$ and $D$ is a Dirac operator, i.e. a first order elliptic
differential operator, then the action functional for Yang-Mills theory is
given by
\be
I_{YM}=Tr_\omega(\Theta^2 |D|^{-n})\;\; ,\label{YM-ac}
\ee
where $\Theta \in \Omega^2_D{\cal A}$ is the curvature 2-form. The role of
$|D|^{-n}\in {\cal L}^{(1,\infty)}$
is to bring $\Theta^2$ into the the ideal $\cL^{(1,\infty)}$. Positivity,
finiteness and covariance are ensured by general properties of the Dixmier
trace.

Let us study the trace theorem (\ref{trateo}) for the Yang-Mills
action (\ref{YM-ac}) in some more detail. We have (assuming $D=|D|$ for
simplicity)
\be
I_{YM} = \lim_{p\ra n+} (p-n)tr (\Theta^2 D^{-p})\;\; .
\ee
The trace, which is finite for $p>n$, can be rewritten as
\be
 (\Theta^2 D^{-p}) = {1\over \Gamma({p \over 2})}
\int_0^{\infty}\!\!dt\, \Theta^2 t^{{p\over2} -1} \exp (-tD^2)\;\; .
\ee
For this expression we can now apply the heat kernel expansion \cite{BGV},
i.e. there is a unique formal solution to the heat equation
\be
(\partial_t + \tilde{\Delta}_x)\bo{k}_t(x,y)=0
\ee
with $\tilde{\Delta}_x = D^2_x$ such that
\be
\int_0^{\infty}\!\!dt\, \Theta^2 t^{{p\over2} -1} \exp(-t\tilde{\Delta_x})\;
{g(x)}^{1\over 4}s(x) =
\int_0^{\infty}\!\!dt\, \Theta^2 t^{{p\over2} -1}
\int \!\!d^{\, n}\!y\sqrt{g(y)} \, \bo{k}_t(x,y)s(y)
\;\; ,\;\; s\in \cH\;\; . \label{hker}
\ee
The kernel $\bo{k}_t$ has the expansion
\be
\bo{k}_t(x,y)=q_t(x,y)\sum_{j=0}^{\infty}t^j \bo{\Phi}_j(x,y)
\ee
where $q_t(x,y)= {g(x)}^{1\over 4}
(\pi t)^{-{n\over 2}}\exp(-{(x-y)^2\over 4t})$ denotes the
Euclidean heat kernel of flat space ($\sqrt{g}$ denotes the canonical density
associated to the Riemannean metric on $M$) and
the $\bo{\Phi}_j$'s can be computed recursively. One finds \cite{BGV} that
\be
\bo{\Phi}_0(x,x) = \bo{id}_{End({\cal H})} \;\; ,\;\; \bo{\Phi}_1(x,x) =
{\textstyle {1\over 6}} \bo{R} -\bo{F}\; ,\ldots \;\;\; .
\ee
where $\bo{R}=R\cdot \bo{id}_{End({\cal H})}$, $R$ is the curvature scalar and
$\bo{F}\in \mbox{End}({\cal H})$ is determined by $\tilde{\Delta}$, i.e. $D$.

Now we can represent $\Theta^2D^{-p}$ by the following expansion
\be
\Theta^2 D^{-p}_xs(x) = (4\pi)^{-{ n\over 2}}\Theta^2\sum_{j=0}^{\infty}
\int_0^{\infty}\!\! dt\,t^{{ p-n\over 2}+j-1}\int\!\!d^{\, n}y\sqrt{g(y)}
\bo{\Phi}_j(x,y)\exp(-{\textstyle{d^2(x,y)\over 4t}})s(y)
\ee
If one takes the trace of this expansion and considers the limit $p\ra n$
the first term in the expansion becomes singular and contributes to
the residue. Therefore eq.(\ref{YM-ac}) can be written as
\be
I_{YM} =\int\!\!d^{\, n}x \sqrt{g}\, tr(\Theta^2)
\ee
However, if we would take the limit $p \ra n-2$ the second term in the
expansion would evolve the same singularity as the first term does in
the case $p\ra n$. If we could pick out this term we would obtain an
functional which contains the curvature scalar and hence it would be a
good candidate for a gravity action. Of course, the first term is in the
limit $p\ra n-2$ horribly divergent and therefore the whole procedure is
ill defined. Thus we have to use a different tool to extract this coefficient.
Fortunatly there is a unique extension of the Dixmier trace to a larger
class of pseudo-differential operators, the Wodzicki residue \cite{wodz}.
This residue will allow us to compute $\mbox{Res}(D^{-n+2})$. The Wodzicki
residue was introduced to non-commutative geometry by A.~Connes in \cite{co}
and used in \cite{co2} to derive a generalized Polyakov action where it was
also
claimed that $\mbox{Res}(D^{-n+2})$ leads to gravity.
In the following section we will show that for $n \geq 4$, $n$ even,
$\mbox{Res}(D^{-n+2})$ picks out the second coefficient of the heat kernel
expansion of $D^2$.
\section{Symbols of an inverse Laplacian}
 In this section we will prove a relation between the symbols of an inverse
 generalized Laplacian and some intrinsic geometric quantities. This relation
 will be used later to build gravity actions out of generalized Dirac operators
 which have in common, that their square is a generalized Laplacian. First we
 shall introduce some notation and review briefly a few basic properties of
 pseudo-differential operators (see also \cite{gil}).

 In the following $M$ is a compact $n$-dimensional Riemannian manifold and
 $\cal E$ and ${\cal E}^\prime$ are (complex) vector bundles of rank $r$ and
 $s$ on $M$. An $m$-th order differential operator
 $\, L :\Gamma (M,{\cal E})\to\Gamma (M,{\cal E}^\prime )\,$ acting on sections
 of $\cal E$ may be written in local coordinates for suitable trivialisations
 of $\cal E$ as
\be
 (Lu(x))_i=\sum^r_{j=1}\sum^m_{|\alpha |=0}(-\, i)^{|\alpha |}\;
 a^{ij}_\alpha (x)\;\partial^\alpha_x u_j(x)\qquad\forall\; i=1,\dots ,s\; ,
\ee
 where $\alpha =(\alpha_1,\dots ,\alpha_n),\; \alpha_i\in\N_{\!\! 0}$ is a
 multiindex with $|\alpha |=\sum^n_{i=1}\alpha_i$, $a^{ij}$ is a
 $r\times s$-matrix and $\partial^\alpha_x:=
 \frac{\partial^{\alpha_1}}{\partial x^{\alpha_1}_1}\cdots
 \frac{\partial^{\alpha_n}}{\partial x^{\alpha_n}_n}$. Motivated by the
 Fourier representation (on $\R^{\!\! n}$)
\be
 \begin{array}{rl}
   u(x)\!
  &{\displaystyle =(2\pi )^{-\frac{n}{2}}\int_{R^n}d\xi\; e^{i<x,\xi >}\;
   \hat u(\xi )}                                                           \\
  &                                                                        \\
   (Lu)(x)\!
  &{\displaystyle =(2\pi )^{-\frac{n}{2}}\int_{R^n}d\xi\; e^{i<x,\xi >}\;
   \sigma^L (x,\xi)\;\hat u(\xi )}                                          \\
\end{array}
\ee
 one introduces the symbol $\sigma^L(x,\xi )$ associated to $L$ by
\be
 \sigma^L(x,\xi )=\sum^m_{k=0}\sigma^L_k(x,\xi ):=\sum^m_{|\alpha |=0}a_\alpha
 (x)\,\xi^\alpha\; ,
\ee
 with $\xi =\xi_\mu dx^\mu$ and $(x,\xi )\in T^\ast M$. The leading term
 $\sigma^L_m(x,\xi )$ is called the principal symbol of $L$. A differential
 operator $L$ is called elliptic, if its principal symbol                    \\
 $\sigma^L_m(x,\xi )\in\Gamma (T^\ast M,\pi^\ast\hbox{End}({\cal E}))\,$ is
 invertible over the open set $\{ (x,\xi )|\,\xi\ne 0\}$. A ge\-ne\-ra\-lized
 Laplacian $\tilde{\Delta}$ on $\cal E$ is a second order elliptic
 differential operator, such that $\,\sigma^{\tilde{\Delta}}_2=\|\xi \|^2
 \cdot\bo{id}_{End({\cal E})}\,$, or equivalently $\tilde{\Delta}$ is given in
 any system of local coordinates by the expression
\be\label{lasym}
 \tilde{\Delta}=-\,\bo{g}^{\mu\nu}\partial_\mu\partial_\nu +\bo{B}^\mu
 \partial_\mu +\bo{C}\; ,
\ee
 where $\,\bo{g}^{\mu\nu}=g^{\mu\nu}\cdot\bo{id}_{End({\cal E})}\,$ and
 $\,\bo{B}(x),\bo{C} (x)\in\hbox{End}({\cal E})\,$. For later purposes it is
 important to notice here (see \cite{BGV}), that given any generalized
 Laplacian $\,\tilde{\Delta}\,$ on $\,\cal E\,$, there exists a connection
 $\,\bo{\nabla}^{\cal E}\,$ on $\cal E$ and a section $\bo{F}$ of the bundle
 $\,\hbox{End}({\cal E})\,$, such that $\tilde{\Delta}$ can be written as
\be\label{lap}
 \tilde{\Delta}=-\, g^{\mu\nu}(\bo{\nabla}^{\cal E}_\mu
 \bo{\nabla}^{\cal E}_\nu -\Gamma^\rho_{\mu\nu}\bo{\nabla}^{\cal E}_\rho )+
 \bo{F} :=\Delta^{\!\!\nabla} +\bo{F}\; ,
\ee
 with $\,\Gamma^\rho_{\mu\nu}\,$ the components of the Levy-Civita connection
 and $\,\Delta^{\!\!\nabla} u=-\, tr(\bo{\nabla}^{T^\ast M\otimes{\cal E}}
 \bo{\nabla}^{\cal E}u)\,$ the Laplacian corresponding to the connection
 $\bo{\nabla}^{\cal E}$ where trace denotes contraction with the metric
 $g\in\Gamma (M,TM\otimes TM)$. From differential operators one comes to
 pseudo-differential operators ($\Psi DO$s for short) by enlarging the space
 of symbols. An appropriate symbol-space for our purposes is the space
 $S^m(U)$ defined by :                                                       \\
 Let $S^m(U)$ ($m\in\R ,\; U\subset\R^{\!\! n}$) be the space of functions
 $\sigma (x,\xi )$ with $(x,\xi )\in K\times\R^{\!\! n}$ and $K$ a compact
 subset of $U$, which satisfy the condition
\be
 |\partial^\alpha_x\partial^\beta_\xi\;\sigma (x,\xi )|\le\, c_{\alpha\beta}
 (1+\|\xi\| )^{m-|\beta |}\qquad\forall\, (x,\xi )\in K\times\R^{\!\! n}\; ,
\ee
 where $c_{\alpha\beta}$ is a constant. We will consider $\Psi DO$s where the
 matrix entries of $\sigma (x,\xi )$ belong to $S^m (U)$. Many of the main
 properties remain true when passing from differential operators to $\Psi DO$s.
 Any $\Psi DO$ $P$ may be defined by a complete symbol which has an asymtotic
 expansion $\,\sigma^P(x,\xi )\sim\sum^\infty_{k=0}\sigma^P_{m-k}(x,\xi )\,$,
 where now $m$ can be any real number, and the $\sigma_{m-k}(x,\xi )$ are
 matrices of smooth functions, which are still homogenous in $\xi$ of degree
 $(m-k)$. The sign $\sim$ denotes equivalent modulo infinitely smoothing
 operators. The symbol of the composition of two $\Psi DO$s $P_1$ and $P_2$ is
 given by the 'Leibnitz rule'
\be \label{leib}
 \sigma^{P_1\circ P_2}(x,\xi )\sim\sum^\infty_{|\alpha |=0}(-\, i)^{|\alpha |}
 \;\frac{1}{\alpha !}\;\partial^\alpha_\xi\sigma^{P_1}\;\partial^\alpha_x
 \sigma^{P_2}
\ee
 with $\,\alpha !=\alpha_1!\cdots\alpha_n!\,$. The last fact that we have to
 state here, is that there exists a unique trace (with certain properties) on
 the algebra of $\Psi DO$s, called the Wodzicki residue \cite{wodz}. For a
 $\Psi DO\,$ $P$, acting on sections of a vector bundle $\cal E$ over a compact
 Riemannian manifold $M$, this is defined by
\be
 Res(P):=\frac{\Gamma (\frac{n}{2})}{2\,\pi^{\frac{n}{2}}}\,
 \int_{S^\ast M}\; tr\, (\sigma^P_{-n}(x,\xi ))\; ,
\ee
 with $\, S^\ast M=\{ (x,\xi )\in T^\ast M\, |\;\|\xi\|^2=1\}\,$ the cosphere
 bundle on $M$. The coefficient in front of the integral is the
 normalisation\footnote{Other authors may use different normalisations.} of the
 volume of $S^{n-1}$. Now we have the tools to discuss another version of
 A.~Connes trace theorem \cite{co} :                                         \\
 For $M$ a compact $n$ dimensional Riemannian manifold and $P$ a $\Psi DO$ of
 order $-\, n$ acting on sections of a (complex) vector bundle $\cal E$ on $M$
 the following relation holds
\be\label{cotra}
 TR_\omega (P)=Res(P)
\ee
 Moreover $Res(P)$ only depends on the conformal class of the metric.        \\
 For a proof we also refer to \cite{VaBo}. In an important work \cite{wodz}
 M.~Wodzicki has shown, that the residue is the unique extension of the
 Dixmier trace to $\Psi DO$s which are not in $\,{\cal L}^{(1,\infty )}
 (\Gamma_{L^2}(M,{\cal E}))\,$. We will use this fact in the following theorem,
 which is the main result of our article, where the relevant $\Psi DO$ is not
 of order $-\, dim\, M$.

 {\it{\bf Theorem} For $M$ a compact $n$ dimensional ($n\ge 4$, even)
 Riemannian manifold and $\tilde{\Delta}$ a generalized Laplacian acting on
 sections of a (complex) vector bundle $\cal E$ on $M$ the following relation
 holds
\be
 Res({\tilde{\Delta}}^{-1})={\textstyle\frac{n-1}{2}}\,\int_M\! d^{\, 4}x\,
 \sqrt{g}\;\, tr\, (\bo{\Phi}_1(x,x,\tilde{\Delta}))\; ,
\ee
 where on the right-hand side
 $\;\bo{\Phi}_1(x,x,\tilde{\Delta})=\frac{1}{6}\,\bo{R} -\bo{F}\;$ is the
 diagonal part of the second coefficient of the heat kernel expansion of
 $\tilde{\Delta}\,$.}

 {\bf Remark :} One important observation (see M.~Wodzicki \cite{wodz}
 proposition 7.11 and remark 7.13) is, that $\,\int_{S^{n-1}}\! d\xi
 d^{\, n}x\;\sqrt{g}\; tr\, (\sigma^P_{-n}(x,\xi ))\,$ is a scalar density,
 even though symbols which are not principal symbols are in general not
 covariant geometric quantities.

 {\bf Proof :}                                                               \\
 The proof will be established in three steps
\begin{iliste}
 \item Calculation of $\,\sigma^{{\tilde{\Delta}}^{-\frac{n}{2}+1}}_{-n}
  (x,\xi )\,$ by a parametrix construction.
 \item Integration over the cosphere bundle (using normal coordinates).
 \item Converting the result into geometric quantities.
\end{iliste}
 (i) Let $\;\sigma^{\tilde{\Delta}}(x,\xi ):=\sigma_2+\sigma_1+\sigma_0\;$
 with $\;\sigma_2\;$ proportional to $\;\bo{id}_{End({\cal E})}=:\bo{1}\,$.
 Introduce a new $\Psi DO$ $P$ by $\;\sigma^P(x,\xi )=\sigma^P_{-2}:=
 (\sigma_2)^{-1}$. According to the composition rule (\ref{leib}) we have
\be\label{para}
 \begin{array}{rcl}
    \sigma^{\tilde{\Delta}\circ P-1}
  &\sim
  &{\displaystyle\sum^\infty_{|\alpha |=0}(-\, i)^{|\alpha |}\;
   \frac{1}{\alpha !}\;\partial^\alpha_\xi\sigma^{\tilde{\Delta}}\;
   \partial^\alpha_x\sigma^{-1}_2-\bo{1}}                                    \\
  &&                                                                         \\
  &=
  &{\displaystyle\sum^2_{k=1}\sum^k_{|\alpha |=0}(-\, i)^{|\alpha |}\;
   \frac{1}{\alpha !}\,\partial^\alpha_\xi\sigma_{|\alpha |+2-k}\;
   \partial^\alpha_x\sigma^{-1}_2}                                           \\
  &&                                                                         \\
  &:=
  &{\displaystyle -\, r(x,\xi )}                                             \\
 \end{array}
\ee
 With the notation $\;\sigma^{P_1}\circ\sigma^{P_2}:=\sigma^{P_1\circ P_2}\;$
 relation (\ref{para}) leads to $\sigma^{\tilde{\Delta}}\circ (\sigma^P\circ
 (\bo{1} -r)^{-1})\sim\bo{1}$. Using the geometric series in symbol-space
 (this can be done because $r$ is of order $-1$) one
 obtains
\be
 \sigma^{{\tilde{\Delta}}^{-1}}(x,\xi )\sim\;\sigma^{-1}_2\circ
 \sum^\infty_{k=0}r^{\circ k}\; .
\ee
 We begin to compute
\be
 \begin{array}{rcl}
   r_{-k}(x,\xi )
  &=
  &\!{\displaystyle \sum^k_{|\alpha |=0}(-\, i)^{|\alpha |}\;
   \frac{1}{\alpha !}\;\partial^\alpha_\xi\sigma_{|\alpha |+2-k}\;
   \partial^\alpha_x\sigma^{-1}_2\; ,}                                       \\
  &&                                                                         \\
   r_{-1}(x,\xi )
  &=
  &\!{\displaystyle -\,\sigma^{-1}_2\sigma_1-i\,\sigma^{-2}_2\;
   \partial_{\xi_\mu}\sigma_2\;\partial_{x^\mu}\sigma_2\; ,}                 \\
  &&                                                                         \\
   r_{-2}(x,\xi )
  &=
  &\!{\displaystyle -\,\sigma^{-1}_2\sigma_0-\sigma^{-2}_2(i\,
   \partial_{\xi_\mu}\sigma_1\;\partial_{x^\mu}\sigma_2+
   {\textstyle\frac{1}{2}}\;\partial_{\xi_\mu}\partial_{\xi_\nu}\sigma_2\;
   \partial_{x^\mu}\partial_{x^\nu}\sigma_2)}                                \\
  &&                                                                         \\
  &
  &\!{\displaystyle +\,\sigma^{-3}_2\;\partial_{\xi_\mu}\partial_{\xi_\nu}
   \sigma_2\;\partial_{x^\mu}\sigma_2\;\partial_{x^\nu}\sigma_2\; ,}         \\
  &&                                                                         \\
   r_{-k}(x,\xi )
  &=
  &\! 0\qquad\qquad\forall\; k>2\; .
 \end{array}
\ee
 Again by relation (\ref{leib}) we further have
\be
 \sum^\infty_{k=0}r^{\circ k}=\sum^\infty_{j=0}s_{-j}\qquad\hbox{with}\qquad
 s_0=\bo{1}\, ,\quad s_{-1}=r_{-1}\, ,\quad s_{-2}=r^2_{-1}+r_{-2}\, ,\quad
 \dots\; .
\ee
 From this we can read off the symbol of $\,{\tilde{\Delta}}^{-1}\,$ :
\be
 \sigma^{{\tilde{\Delta}}^{-1}}(x,\xi)\sim\sum^\infty_{l=2}
 \sigma^{{\tilde{\Delta}}^{-1}}_{-l}\quad\hbox{with}\quad
 \sigma^{{\tilde{\Delta}}^{-1}}_{-l}(x,\xi )=\sum^{l-2}_{|\alpha |=0}
 (-\, i)^{|\alpha |}\;\frac{1}{\alpha !}\;\partial^\alpha_\xi
 \sigma^{-1}_2\;\partial^\alpha_xs_{|\alpha |+2-l}\; .
\ee
 We will only need the first three non-vanishing terms :
\be
 \begin{array}{rcl}
   \sigma^{{\tilde{\Delta}}^{-1}}_{-2}(x,\xi )
  &=
  &\!{\displaystyle\sigma^{-1}_2\; ,\qquad
   \sigma^{{\tilde{\Delta}}^{-1}}_{-3}(x,\xi )=\sigma^{-1}_2\, r_{-1}}\; ,   \\
  &&                                                                         \\
   \sigma^{{\tilde{\Delta}}^{-1}}_{-4}(x,\xi )
  &=
  &\!{\displaystyle \sigma^{-1}_2\, (r^2_{-1}+r_{-2})+\,i\,\sigma^{-2}_2\;
   \partial_{\xi_\mu}\sigma_2\;\partial_{x^\mu}r_{-1}\; .}                   \\
 \end{array}
\ee
 More generally we get
\be
 \sigma^{{\tilde{\Delta}}^{-m}}(x,\xi)\sim\,\sigma^{{\tilde{\Delta}}^{-m+1}}
 \circ\sigma^{{\tilde{\Delta}}^{-1}}\sim\sum^\infty_{|\alpha |=0}
 (-\, i)^{|\alpha |}\;\frac{1}{\alpha !}\;\partial^\alpha_\xi
 \sigma^{{\tilde{\Delta}}^{-m+1}}\;\partial^\alpha_x
 \sigma^{{\tilde{\Delta}}^{-1}}:=\sum^\infty_{l=2m}
 \sigma^{{\tilde{\Delta}}^{-m}}_{-l}\; ,
\ee
\be\label{masterfo}
 \hbox{with}\qquad\sigma^{{\tilde{\Delta}}^{-m}}_{-l}(x,\xi )=
 \sum^{l-2m}_{|\alpha |=0}\quad\sum^{2+l-|\alpha |-2m}_{k=2}
 (-\, i)^{|\alpha |}\;\frac{1}{\alpha !}\;\partial^\alpha_\xi
 \sigma^{{\tilde{\Delta}}^{-m+1}}_{|\alpha |+k-l}\;\partial^\alpha_x
 \sigma^{{\tilde{\Delta}}^{-1}}_{-k}\; .\qquad\hbox{}
\ee
 Using this and $\,\sigma^{{\tilde{\Delta}}^{-m}}_{-2m}\equiv\sigma^{-m}_2\,$
 we get the recursion relations
\be\label{recure}
 \begin{array}{rcl}
   \sigma^{{\tilde{\Delta}}^{-\frac{n}{2}+1}}_{-n}(x,\xi )
  &=
  &\!{\displaystyle\sum^2_{|\alpha |=0}\;\sum^{4-|\alpha|}_{k=2}
   (-\, i)^{|\alpha |}\;\frac{1}{\alpha !}\;\partial^\alpha_\xi
   \sigma^{{\tilde{\Delta}}^{-\frac{1}{2}+2}}_{|\alpha |+k-n}\;
   \partial^\alpha_x\sigma^{{\tilde{\Delta}}^{-1}}_{-k}}                     \\
  &&                                                                         \\
  &=
  &\!{\displaystyle\sigma^{{\tilde{\Delta}}^{-\frac{n}{2}+2}}_{2-n}\,
   \sigma^{-1}_2+\sigma^{{\tilde{\Delta}}^{-\frac{n}{2}+2}}_{3-n}\,
   \sigma^{{\tilde{\Delta}}^{-1}}_{-3}+
   \sigma^{{\tilde{\Delta}}^{-\frac{n}{2}+2}}_{4-n}\,
   \sigma^{{\tilde{\Delta}}^{-1}}_{-4}-\, i\,\partial_{\xi_\mu}
   \sigma^{{\tilde{\Delta}}^{-\frac{n}{2}+2}}_{3-n}\;\partial_{x^\mu}
   \sigma^{-1}_2}                                                            \\
  &&                                                                         \\
  &&\!{\displaystyle -\, i\,\partial_{\xi_\mu}\sigma^{-\frac{n}{2}+2}_2\;
   \partial_{x^\mu}\sigma^{{\tilde{\Delta}}^{-\frac{n}{2}+2}}_{-3}
   -\,{\textstyle\frac{1}{2}}\;\partial_{\xi_\mu}\partial_{\xi_\nu}
   \sigma^{-\frac{n}{2}+2}_2\;\partial_{x^\mu}\partial_{x^\nu}\sigma^{-1}_2} \\
 \end{array}
\ee
 and again with relation (\ref{masterfo})
\be\label{recurz}
 \sigma^{{\tilde{\Delta}}^{-\frac{n}{2}+2}}_{3-n}(x,\xi)=
 \sigma^{{\tilde{\Delta}}^{-\frac{n}{2}+3}}_{5-n}\,\sigma^{-1}_2+
 \sigma^{-\frac{n}{2}+3}_2\,\sigma^{{\tilde{\Delta}}^{-1}}_{-3}-\, i\,
 \partial_{\xi_\mu}\sigma^{-\frac{n}{2}+3}_2\;\partial_{x^\mu}
 \sigma^{-1}_2\; .
\ee
 Now we could proceed in calculating $\, Res\, ({\tilde{\Delta}}^{-1})\,$ in
 an arbitrary coordinate system, by solving the recursion formulas, but some
 terms would then become rather clumsy, so it is more convenient to pass to
 Riemannian normal coordinates at this stage.

 (ii) The Taylor expansion of the function $g^{\mu\nu}$ in Riemannian normal
 coordinates ${\bf X}$ around a point $x_0$ up to order $\, {\bf X}^3\,$
 reads :
\be\label{normal}
 g^{\mu\nu}({\bf X})=\delta^{\mu\nu}-{\textstyle\frac{1}{3}}\,
 R^{\mu\phantom{\rho}\nu}_{\phantom{\mu}\rho\phantom{\nu}\lambda}(x_0)\;
 {\bf X}^\rho{\bf X}^\lambda +O({\bf X}^3)\; .
\ee
 From now on we will calculate everything with respect to these coordinates :
\be
 \begin{array}{rcl}
   r_{-1}(x_0,\xi )
  &=
  &{\displaystyle -\,\sigma^{-1}_2\,\sigma_1\; ,\qquad r_{-2}(x_0,\xi )=
   -\,\sigma^{-1}_2\,\sigma_0+{\textstyle\frac{2}{3}}\,\sigma^{-2}_2
   \delta^{\rho\sigma}\,
   R^{\mu\phantom{\rho}\nu}_{\phantom{\mu}\rho\phantom{\nu}\sigma}\,\xi_\mu
   \xi_\nu\; ,}                                                              \\
  &&                                                                         \\
   \sigma^{{\tilde{\Delta}}^{-1}}_{-2}(x_0,\xi )
  &=
  &{\displaystyle\sigma^{-1}_2\; ,\qquad\sigma^{{\tilde{\Delta}}^{-1}}_{-3}
   (x_0,\xi )=-\,\sigma^{-2}_2\,\sigma_1\; ,}                                \\
  &&                                                                         \\
   \sigma^{{\tilde{\Delta}}^{-1}}_{-4}(x_0,\xi )
  &=
  &{\displaystyle -\,\sigma^{-2}_2\,\sigma_0+\sigma^{-3}_2\left(\sigma^2_1-\,
   2\, i\,\delta^{\rho\mu}\;\partial_{x^\rho}\sigma_1\;\xi_\mu +\,
   {\textstyle\frac{2}{3}}\,\delta^{\rho\sigma}\,
   R^{\mu\phantom{\rho}\nu}_{\phantom{\mu}\rho\phantom{\nu}\sigma}\;
   \xi_\mu\xi_\nu\right)\; .}                                                \\
 \end{array}
\ee
 With this we can easily solve the recursion relation (\ref{recurz})
\be\label{recurd}
 \begin{array}{rl}
   \sigma^{{\tilde{\Delta}}^{-\frac{n}{2}+2}}_{3-n}(x_0,\xi )
  &{\displaystyle =\sigma^{{\tilde{\delta}}^{-\frac{n}{2}+3}}_{5-n}\,
   \sigma^{-1}_2+\sigma^{-\frac{n}{2}+3}_2\,
   \sigma^{{\tilde{\Delta}}^{-1}}_{-3}}                                      \\
  &                                                                          \\
   \leadsto\quad
  &{\displaystyle\sigma^{{\tilde{\Delta}}^{-\frac{n}{2}+2}}_{3-n}(x_0,\xi)=
   ({\textstyle\frac{n}{2}}-2)\,\sigma^{-\frac{n}{2}+2}\, r_{-1}\; .}\qquad  \\
 \end{array}
\ee
 Inserting eq. (\ref{recurd}) in relation (\ref{recure}) yields
\be
 \begin{array}{rl}
   \sigma^{{\tilde{\Delta}}^{-\frac{n}{2}+1}}_{-n}(x_0,\xi )\!
  &{\displaystyle =\sigma^{{\tilde{\Delta}}^{-\frac{n}{2}+2}}_{2-n}\,
   \sigma^{-1}_2+({\textstyle\frac{n}{2}}-1)\,\sigma^{-\frac{n}{2}+2}_2
   \sigma^{{\tilde{\Delta}}^{-1}}_{-4}+({\textstyle\frac{n}{2}}-2)\,
   \sigma^{-\frac{n}{2}}_2\sigma_0}                                          \\
  &                                                                          \\
   \leadsto\quad
  &{\displaystyle \sigma^{{\tilde{\Delta}}^{-\frac{n}{2}+1}}_{-n}(x_0,\xi )
   ={\textstyle\frac{n-2}{8}}\left( n\,\sigma^{-\frac{n}{2}+2}_2
   \sigma^{{\tilde{\Delta}}^{-1}}_{-4}+(n-4)\,\sigma^{-\frac{n}{2}}_2\,
   \sigma_0\right)\; .}                                                      \\
 \end{array}
\ee
 With the help of the following identity
\be
 \int_{S^{n-1}}\! d\xi\; A^{\mu\nu}\;\xi_\mu\xi_\nu =
 \frac{2\,\pi^{\frac{n}{2}}}{n\,\Gamma (\frac{n}{2})}\; g_{\mu\nu}\,
 A^{\mu\nu}\; ,
\ee
 and using the explicit symbol $\,\sigma^{\tilde{\Delta}}=\bo{g}^{\mu\nu}\,
 \xi_\mu\xi_\nu +i\,\bo{B}^\mu\xi_\mu +\bo{C}\,$ (see eq. (\ref{lasym})) we get
\be
 \int_{S^{n-1}}\! d\xi\;\sigma^{{\tilde{\Delta}}^{-1}}_{-4}(x_0,\xi )=
 \frac{2\,\pi^{\frac{n}{2}}}{(\frac{n}{2}-1)!}\left(-\,\bo{C}+
 {\textstyle\frac{4}{n}}({\textstyle\frac{1}{6}}\,\bo{R}+\,
 {\textstyle\frac{1}{2}}\,\partial_\mu\bo{B}^\mu |_{X=0}-
 {\textstyle\frac{1}{4}}\,\bo{B}_\mu\,\bo{B}^\mu )\right)\; ,
\ee
 and therefore
\be
 \int_{S^{n-1}}\! d\xi\;\sigma^{{\tilde{\Delta}}^{-\frac{n}{2}+1}}_{-n}
 (x_0,\xi )=\frac{2\,\pi^{\frac{n}{2}}}{(\frac{n}{2}-2)!}
 \left({\textstyle\frac{1}{6}}\,\bo{R}-\bo{C}+{\textstyle\frac{1}{2}}\,
 \partial_{x^\mu}\bo{B}^\mu |_{X=0}-{\textstyle\frac{1}{4}}\,
 \bo{B}_\mu\bo{B}^\mu\right)\; .
\ee

 (iii) Comparison of eq. (\ref{lasym}) with eq. (\ref{lap}), together with the
  definition $\,\bo{\nabla}^{\cal E}_\mu :=\partial_\mu +\bo{\Omega}_\mu\,$
 leads to
\be
 \begin{array}{rl}
   \bo{B}^\mu =
  &\!\!\bo{g}^{\rho\nu}\Gamma^\mu_{\rho\nu}-2\, g^{\mu\nu}\bo{\Omega}_\nu    \\
  &                                                                          \\
   \bo{C} =
  &\!\! -\, g^{\mu\nu}(\partial_\mu\bo{\Omega}_\nu -\Gamma^\rho_{\mu\nu}
   \bo{\Omega}_\rho +\bo{\Omega}_\mu\bo{\Omega}_\nu )+\bo{F}\; .
 \end{array}
\ee
 Passing to normal coordinates we find
\be
 \bo{F} (x_0)=\bo{C}-{\textstyle\frac{1}{2}}\,\partial_\mu\bo{B}^\mu |_{X=0}+
 {\textstyle\frac{1}{4}}\bo{B}_\mu\bo{B}^\mu\; ,
\ee
 and so we finally arrive at
\be\label{result}
 \int_{S^{n-1}}\! d\xi\;\sigma^{{\tilde{\Delta}}^{-\frac{n}{2}+1}}_{-n}
 (x_0,\xi )=\frac{2\,\pi^{\frac{n}{2}}}{(\frac{n}{2}-2)!}
 \left({\textstyle\frac{1}{6}}\,\bo{R}-\bo{F}\right)\; .
\ee
 However, as already mentioned above in the remark, the left-hand side of this
 equation is a scalar density and therefore, because the right-hand side is a
 covariant quantity, eq. (\ref{result}) holds in any system of local
 coordinates.$\hbox{}\hfill\Box$                                             \\
 With this theorem  we now have a tool to construct gravity actions  just by
 choosing a Dirac operator $D$, squaring it and read off $\,\bo{F} (D^2)\,$.
 This will be the topic of the next section.

\section{Dirac Operators and Gravity Actions}
 In this section we will consider three different types of Dirac operators and
 the gravity actions associated to them via the main theorem of sec.4. For
 gravity actions on manifolds satisfying the same hypotheses as in the theorem
 we take
\be\label{gr-ac}
 I_{GR}:=\, -\,{\textstyle\frac{2}{r\, (n-1)}}\, Res\, (D^{-n+2})
 =\int_M\! d^{\, n}x\;\sqrt{g}\,\left(-{\textstyle\frac{1}{6}}\, R+
 {\textstyle\frac{1}{r}}\, tr\, (\bo{F})\right)\; ,
\ee
 with $r$ the rank of the involved vector bundle $\cal E$ and $R$ the
 scalar curvature of $M$. By a Dirac operator $D$ we understand an odd
 first-order elliptic differential operator acting on sections of a
 $\Z_{\!\! 2}$-graded vector bundle ${\cal E}={\cal E}^+\oplus{\cal E}^-$
\be
 D\; :\;\Gamma (M,{\cal E}^\pm )\,\to\,\Gamma (M,{\cal E}^\mp ),
\ee
 such that $D^2$ is a generalized Laplacian. For these vector bundles we take
 Clifford modules over an even-dimensional Riemannian manifold $M$. Such a
 Clifford module is a $\,\Z_{\!\! 2}$-graded bundle $\cal E$ on $M$ with a
 graded action ('Clifford-action') of the Clifford bundle $\,{\cal{CL}}(M)=
 {\cal{CL}}^+\oplus{\cal{CL}}^-\,$ on it :
\be
 {\cal{CL}}^+\cdot{\cal E}^\pm\subset{\cal E}^\pm\; ,\qquad
 {\cal{CL}}^-\cdot{\cal E}^\pm\subset{\cal E}^\mp\; .
\ee
 With this data the Dirac-K-cycle reads $\, ({\cal A},{\cal H},D)\equiv
 (C^\infty (M),\Gamma_{L^2}(M,{\cal E}),D)\,$. Probably the easiest example
 of this setting is the Dirac operator $\, D=d+d^\ast\,$ (called the
 signature operator if $dim\, M$ is divisible by four) acting on the exterior
 bundle $\,\Lambda T^\ast M\,$, here considered as a Clifford module. To this
 Dirac operator corresponds the Laplace-Beltrami operator
 $\,\Delta =D^2=dd^\ast +d^\ast d\,$, which is the canonical Laplacian
 associated to $\,\Lambda T^\ast M$ with $\,\bo{F}=0\,$. For this operator the
 action is easily calculated to be $\, I_{GR}=-{\textstyle\frac{1}{6}}
 \int_Md^{\, n}x\,\sqrt{g}\; R\,$. Another quite important Clifford module is
 the spinor bundle over a spin manifold. This is the relevant example, if one
 wants to have fermions in the respective physical model. Actually it is known,
 that for an even-dimensional Riemannian spin manifold any Clifford module is a
 twisted bundle $\,{\cal E}={\cal W}\otimes{\cal S}\,$, where ${\cal W}$ is a
 bundle on which the ${\cal{CL}}$-action is trivial and ${\cal S}$ is the
 spinor bundle. Before going to explicit examples we should mention our
 conventions for the representation matrices $\gamma^a$ of the Euclidean
 Clifford algebra :
\be\label{cliff}
 \{\gamma^a,\gamma^b\} =-\, 2\,\delta^{ab}\,\bo{1}\; ,\quad [\gamma^a,\gamma^b]
 =2\,\gamma^{ab}\; ,\quad tr\,\gamma^a=0\quad\leadsto\quad tr\,\gamma^{ab}=0
 \; .
\ee
 Further we should remark, that the coordinate base $\{\gamma^\mu
 {\scriptstyle (x)}\}$ can as usual be converted via the vielbeins
 $\varepsilon^a_\mu{\scriptstyle (x)}$ to an orthonormal basis $\{\gamma^a\}$
 by $\,\gamma^\mu =\varepsilon^\mu_a\,\gamma^a$.

 {\bf $\bo{\bullet}$ Dirac operator of a Clifford connection}                \\
 Let $\bo{\nabla}^{{\cal E}}$ be a Clifford connection on a Clifford module
 $\cal E$ over a compact $n$ ($n$ as above) dimensional Riemannian manifold
 $M$. Such a connection is defined (with respect to local coordinates) by the
 relation
\be\label{cliffco}
 [\bo{\nabla}^{{\cal E}}_\mu ,\gamma^\nu ]=-\,\gamma^\rho\,
 \Gamma^\nu_{\rho\mu}\; ,
\ee
 so that in the most general case $\bo{\nabla}^{{\cal E}}_\mu$ may explicitely
 be written as
\be
 \bo{\nabla}^{{\cal E}}_\mu :=\bo{\omega}_\mu +\bo{A}_\mu\; ,
\ee
 with $\,\bo{\omega}_\mu :=\partial_\mu -\frac{1}{4}\,\omega_{\mu ab}
 \gamma^{ab}\,$ the Levy-Civita spin connection and $\bo{A}_\mu =A_\mu\cdot
 \bo{1}\,$. The corresponding Dirac operator $D_\nabla$ associated to
 $\bo{\nabla}^{{\cal E}}$ reads
\be
 D_\nabla =\gamma^\mu\,\bo{\nabla}^{{\cal E}}_\mu =\gamma^a\varepsilon^\mu_a
 \,\bo{\nabla}^{{\cal E}}_\mu\; .
\ee
 With the help of the identity
\be\label{levycom}
 \gamma^{\mu\nu}\, [\bo{\omega}_\mu ,\bo{\omega}_\nu ]={\textstyle\frac{1}{2}}
 \,\bo{R}
\ee
 and (\ref{cliff}), (\ref{cliffco}), we get for $\, D^2_\nabla\,$ the
 Lichnerowicz formula
\be
 \begin{array}{rl}
    D^2_\nabla\!\!
   &={\textstyle\frac{1}{2}}\,\{\gamma^\mu\,\bo{\nabla}^{{\cal E}}_\mu ,
    \gamma^\nu\,\bo{\nabla}^{{\cal E}}_\nu\}                                \\
   &                                                                         \\
   &={\textstyle\frac{1}{2}}\,\{\gamma^\mu ,\gamma^\nu\}\,
    \bo{\nabla}^{{\cal E}}_\mu\bo{\nabla}^{{\cal E}}_\nu +\gamma^\mu\,
    [\bo{\nabla}^{{\cal E}}_\mu ,\gamma^\nu ]\,\bo{\nabla}^{{\cal E}}_\nu +
    {\textstyle\frac{1}{2}}\,\gamma^\mu\gamma^\nu\, [
    \bo{\nabla}^{{\cal E}}_\mu ,\bo{\nabla}^{{\cal E}}_\nu ]                 \\
   &                                                                         \\
   &=-\, g^{\mu\nu}\,\bo{\nabla}^{{\cal E}}_\mu\bo{\nabla}^{{\cal E}}_\nu -
    \gamma^\mu\gamma^\rho\;\Gamma^\nu_{\mu\rho}\,\bo{\nabla}^{{\cal E}}_\nu +
    {\textstyle\frac{1}{2}}\,\gamma^{\mu\nu}\, [\bo{\nabla}^{{\cal E}}_\mu ,
    \bo{\nabla}^{{\cal E}}_\nu ]                                             \\
   &                                                                         \\
   &=-\, g^{\mu\nu}\, (\bo{\nabla}^{{\cal E}}_\mu\bo{\nabla}^{{\cal E}}_\nu -
    \,\Gamma^\rho_{\mu\nu}\,\bo{\nabla}^{{\cal E}}_\rho )+
    {\textstyle\frac{1}{2}}\,\gamma^{\mu\nu}\, [\bo{\omega}_\mu ,
    \bo{\omega}_\nu ]+\gamma^{\mu\nu}\;\partial_\mu\bo{A}_\nu                \\
   &                                                                         \\
   &=\Delta^{\!\!\nabla}+{\textstyle\frac{1}{4}}\,\bo{R}+\bo{F}^{{\cal E}/S}
    \; ,                                                                     \\
 \end{array}
\ee
 with $\,\Delta^{\!\!\nabla}\,$ the Laplacian associated to
 $\,\bo{\nabla}^{{\cal E}}\,$ and $\,\bo{F}^{{\cal E}/S}=\gamma^{\mu\nu}\,
 \partial_\mu\bo{A}_\nu\,$ the so-called twisting curvature (see also
 \cite{BGV}). So we identify for our first example $\;\bo{F}=
 \frac{1}{4}\bo{R}+\bo{F}^{{\cal E}/S}\;$ and the resulting gravity action
 (see (\ref{gr-ac})) reads
\be
 I_{GR}=-{\textstyle\frac{1}{r}}\,\int_M\! d^{\, n}x\;\sqrt{g}\;tr\,
 \left({\textstyle\frac{1}{6}}\,\bo{R}-{\textstyle\frac{1}{4}}\,\bo{R}-
 \bo{F}^{{\cal E}/S}\right)={\textstyle\frac{1}{12}}\int_M\! d^{\,n}x\;
 \sqrt{g}\; R\; .
\ee
 We recognize that this action is proportional to the usual Einstein Hilbert
 action and the $\;\bo{A}_\mu$-part ('the photon') drops out completely.
 Actually any connection associated to an additional internal symmetry does
 not contribute to the action because the relevant terms are of the same type
 as the ones for the twisting curvature and the respective traces factorize.

 {\bf $\bo{\bullet}$ Dirac operator with torsion}                            \\
 Let now $\, D_{\tilde\nabla}\,$ be the Dirac operator defined by the same
 data as in the first example plus an additional torsion term i.e.
\be
 {\bo{\tilde\nabla}^{{\cal E}}_\mu} :=\bo{\nabla}^{{\cal E}}_\mu +\bo{T}_\mu
 \; ,
\ee
 with $\,\bo{T}_\mu :=\frac{1}{4}\, t_{\mu ab}\,\gamma^{ab}=\frac{1}{4}\,
 \varepsilon^c_\mu\, t_{cab}\,\gamma^{ab}\,$ and $\, t_{cab}\,$ totally
 antisymmetric. A calculation analogous to the one above shows
\be
 \begin{array}{rcl}
    D^2_{\tilde\nabla}
   &=
   &\! -\, g^{\mu\nu}\,{\bo{\tilde\nabla}^{\cal E}_\mu}
    {\bo{\tilde\nabla}^{\cal E}_\nu}+\gamma^\mu\, [
    {\bo{\tilde\nabla}^{\cal E}_\mu},\gamma^\nu ]\,
    {\bo{\tilde\nabla}^{\cal E}_\nu}+{\textstyle\frac{1}{2}}\,
    \gamma^\mu\gamma^\nu\, [{\bo{\tilde\nabla}^{{\cal E}}_\mu},
    {\bo{\tilde\nabla}^{\cal E}_\nu}]                                        \\
   &&                                                                        \\
   &=
   &\! -\, g^{\mu\nu}\left(\bo{\nabla}^{{\cal E}}_\mu
    \bo{\nabla}^{{\cal E}}_\nu -\Gamma^\rho_{\mu\nu}\,
    (\bo{\nabla}^{{\cal E}}_\rho +\bo{T}_\rho )+6\,\bo{T}_\mu
    \bo{\nabla}^{{\cal E}}_\nu +[\bo{\nabla}^{{\cal E}}_\mu ,\bo{T}_\nu ]+5
    \,\bo{T}_\mu\bo{T}_\nu\right)                                            \\
   &&                                                                        \\
   &&\! +\,{\textstyle\frac{1}{2}}\,\gamma^{\mu\nu}\, [
    \bo{\nabla}^{{\cal E}}_\mu ,\bo{\nabla}^{{\cal E}}_\nu ]+\gamma^{\mu\nu}
    \, [\bo{\nabla}^{{\cal E}}_\mu +{\textstyle\frac{1}{2}}\,\bo{T}_\mu ,
    \bo{T}_\nu ]\; .                                                         \\
 \end{array}
\ee
 By introducing the connection $\,\bo{\hat\nabla}^{\cal E}_\mu :=
 \bo{\nabla}^{{\cal E}}_\mu +3\,\bo{T}_\mu\,$ and using (\ref{levycom}) this
 may be rewritten as
\be
 \begin{array}{rl}
   D^2_{\tilde\nabla}=\!
  &\Delta^{\!\!\hat\nabla}+{\textstyle\frac{1}{4}}\,\bo{R}+\bo{F}^{{\cal E}/S}
   +2\, g^{\mu\nu}\left( [\bo{\nabla}^{{\cal E}}_\mu ,\bo{T}_\nu ]-
   \Gamma^\rho_{\mu\nu}\,\bo{T}_\rho +2\,\bo{T}_\mu\bo{T}_\nu\right)         \\
  &                                                                          \\
  &+\,\gamma^{\mu\nu}\, [\bo{\nabla}^{{\cal E}}_\mu +{\textstyle\frac{1}{2}}
   \,\bo{T}_\mu ,\bo{T}_\nu ]\; .                                            \\
 \end{array}
\ee
 In this case we find
\be
 \begin{array}{rl}
   tr\, (\bo{F})\!\!
  &{\displaystyle =tr\left( {\textstyle\frac{1}{4}}\,\bo{R}+
   \bo{F}^{{\cal E}/S}+2\, g^{\mu\nu}([\bo{\nabla}^{{\cal E}}_\mu ,\bo{T}_\nu ]
   -\Gamma^\rho_{\mu\nu}\,\bo{T}_\rho +2\,\bo{T}_\mu\bo{T}_\nu )+\,
   \gamma^{\mu\nu}\, [\bo{\nabla}^{{\cal E}}_\mu +{\textstyle\frac{1}{2}}\,
   \bo{T}_\mu ,\bo{T}_\nu ]\right)}                                          \\
  &                                                                          \\
  &{\displaystyle ={\textstyle\frac{r}{4}}\, R+tr\left( 4\, g^{\mu\nu}
   \,\bo{T}_\mu\bo{T}_\nu +{\textstyle\frac{1}{2}}\,\gamma^{\mu\nu}\,
   [\bo{T}_\mu ,\bo{T}_\nu ]+\gamma^{\mu\nu}[\bo{\omega}_\mu ,\bo{T}_\nu
   ]\right)\; .}                                                             \\
 \end{array}
\ee
 Using the torsion constraints for the Levy-Civita spin connection
 $\bo{\omega}_\mu$ one can easily show, that $\,\gamma^{\mu\nu}
 [\bo{\omega}_\mu ,\bo{T}_\nu ]\,$ is a boundary term, which does not
 contribute to the action. For the action we thus get
\be
 I_{GR}=\int_M\! d^{\, 4}x\;\sqrt{g}\,\left({\textstyle\frac{1}{12}}\, R -
 {\textstyle\frac{3}{4}}\, t_{abc}\, t^{abc}\right)\; .
\ee
 The remaining torsion terms also drop out by their equations of motion and we
 get the same result as in the first example. From the derivation it is clear,
 that this would also hold if we would start with a connection which is only
 compatible with the metric, since the functional $Res$ singles out the
 Levy-Civita spin connection.

 {\bf $\bo{\bullet}$ Dirac operator for 'product K-cycles'}                  \\
 As a special case for a product K-cycle we shall now consider the
 generalized Dirac operator associated to a 'non-commutative two-point space'.
 Such spaces where used for a derivation of models of the electroweak
 interactions in non-commutative geometry (see \cite{colo} for example). A
 similar setting in the context of gravity was studied in \cite{chams}. Such a
 two-point space is there given by a four dimensional compact Riemannian spin
 manifold $N$, where $N$ is supposed to be also a principal-$G$-bundle (here
 $G=\Z_{\!\! 2}$) $\pi : N\to M$, over a manifold $M$. Now the relevant
 Clifford module on which the Dirac operator acts is ${\cal E}=
 \pi_\ast{\cal S}\,$ a bundle over $M$ with fiber
 $\, S_y=\oplus_{\pi (x)=y}S_x\,$ and $S_x$ the fiber of the spinor bundle  at
 $x\in N$. In this case the $\,\Z_{\!\! 2}$-equivariant Dirac operator and
 its square have the form
\be
 D=
 \left(
  \begin{array}{cc}
   \gamma^\mu\,\bo{\tilde{\nabla}^{\cal E}_\mu} &\gamma^5\, \phi            \\
                                                &                           \\
    \gamma^5\phi  &\gamma^\mu\,\bo{\tilde{\nabla}^{\cal E}_\mu}             \\
  \end{array}
 \right)
 \quad\leadsto\quad
 D^2=
 \left(
  \begin{array}{cc}
    D^2_{\tilde\nabla}+\phi^2\,\bo{1}
   &\gamma^\mu\gamma^5\,\partial_\mu\phi                                    \\
   &                                                                        \\
    \gamma^\mu\gamma^5\,\partial_\mu\phi
   &D^2_{\tilde\nabla}+\phi^2\,\bo{1}                                       \\
  \end{array}
 \right)
\ee
 with $\,\gamma^5:=\frac{1}{4!}\,\epsilon_{abcd}\,
 \gamma^a\gamma^b\gamma^c\gamma^d\,$. From this and with the knowledge of the
 preceeding examples we can read off the resulting gravity action (with
 the torsion $\bo{T}$ already eliminated by its equations of motion) :
\be
 I_{GR}=\int_M\! d^{\, 4}x\,\sqrt{g}\;\left({\textstyle\frac{1}{12}}\, R+
 \phi^2\right)
\ee

Until now we did not say anything about the nature of $\phi$.
In \cite{chams} the derivation of a gravity action led to kinetic
term for $\phi$, i.e. $\phi$ is a scalar field. However, in
our case there is no kinetic term for $\phi$ and therefore we interpret
the $\phi^2$ term as a cosmological constant. Thus our result is different
from that obtained in \cite{chams}. However, the authors of \cite{chams}
followed a different philosophy which involves the generalized de Rham
algebra $\OD\cA$ whereas in our approach only the Dirac operator of
a K-cycle is used to derive a gravity action.

Now we will show that this result remains true for product K-cycles over
algebras \cA which are a tensor product of the algebra of functions
$C^\infty (M)$ on $M$ and a finite dimensional unital matrix algebra
$\cA_{Mat}$ which in general is a
direct sum of matrix algebras. The number of terms in this sum corresponds
to the number of points of the discrete geometry. The K-cycle over
$C^\infty (M)$ is
given by $(D,\cH_1)$, where $D$ denotes the usual Dirac operator and
$\cH_1$ is the Hilbert-space of square-integrable spin-sections. The second
K-cycle, the K-cycle over $\cA_{Mat}$, is the tuple $(\cM, \cH_2)$ where
$\cH_2$ is a finite dimensional representation space of $\cA_{Mat}$, i.e.
$\cA_{Mat} \subset \mbox{End}(\cH_2)$. The Dirac operator for this finite
dimensional algebra is given by $\cM \in \mbox{End}(\cH_2)$
with \cite{KPPW}
\be
tr (\cM a) = 0\;\; , \;\; \forall a \in \cA\;\; .\label{trco}
\ee
The product K-cycle \cite{cobuch} over the product algebra $\cA$
is the tuple $(\cD , \cH)$ with
\be \begin{array}{rcl}
\cH & = & \cH_1\otimes\cH_2\\
    &   &   \\
\cD & = & D\otimes \bo{1} + \gamma^5 \otimes \cM\;\; .
\end{array}\label{prodcyc}
\ee
One of the main ideas in \cite{chams} was that the geometry of discrete
space may depend on points of the manifold $M$. This means in our case that
we allow for a space dependent $\cM$. Formally the Dirac operator $\cD$ is
now a first-order differential operator on                                 \\
$\cH^\prime=\Gamma(M,\cE)$ where $\cE$ is a Clifford module
with typical fibre $S\otimes \cH_2$, $S$ denotes the spinor
space. The next step is to take the square of $\cD$ and we get
\be
\cD^2 = D^2\otimes \bo{1} + \bo{1}\otimes \cM^2 +
\gamma^5\otimes [\dslash , \cM ] \;\;\; ,\label{Dquad}
\ee
which again is a generalized Laplacian such that we can apply the theorem
of sec.4.
Taking into account condition eq.(\ref{trco}) the gravity action for
this system is given by
\be
I^\prime_{GR} = \int_M \!\! d^{\, 4}x\sqrt{g}\,
({\textstyle {1\over12}}R + {\textstyle {1\over dim {\cal H}_2}}
tr(\cM^2))\;\;\; .\label{np-ac}
\ee
We note that also in this more general `n-point' case there is no kinetic
term for fields contained in the matrix derivative. The reason is that
derivatives of $\cM$, which could lead to a kinetic term in the Lagrangian,
appear only in the third term of eq.(\ref{Dquad}) and therefore cannot
contribute to the Lagrangian because of eq.(\ref{trco}). As for the
two-point case we interpret the term $tr(\cM^2)$ in eq.(\ref{np-ac})
as a cosmological constant. Another consequence of the vanishing of
the kinetic term for $\cM$ is that the product K-cycle (\ref{prodcyc})
is sufficient to describe gravity for continuous space-time $\times$
discrete space.
\section{Conclusions}
We have shown that an action for gravity can be obtained in the
framework of non-commutative geometry via the Wodzicki residue. We
proved that $\mbox{Res} (D^{-n+2})$ picks out the second coefficient
of the heat kernel expansion of $D^{2}$ where $D$ is a Dirac
operator on an $n$ dimensional compact Riemannian manifold with $n\geq 4$ and
$n$ even. In this article we applied our result to a conventional
geometric set up where we could check that this procedure leads to
usual gravity, i.e. it leads to the Einstein Hilbert action. We also
considered an extension of commutative geometry to
non-commutative geometry given by the tensor product of the algebra of smooth
functions on a manifold and an finite dimensional matrix algebra. In this
case we obtained gravity with a cosmological constant.

A natural furhter question is how to couple Yang-Mills fields to gravity.
As we have seen, a Yang-Mills connection which can be part of the
Dirac operator does not contribute to the Wodzicki residue that leads to
an action of gravity. A solution to this problem is
given by adding the Yang-Mills action given by eq.(\ref{YM-ac}) to the
gravity action eq.(\ref{gr-ac}).

However, there are some limitations to our approach. The first, common
to all models in non-commutative geometry which use Dixmier trace resp.
Wodzicki residue, is the fact that so far we
can only describe Riemannian geometry but not pseudo-Riemannian space-time.
Furthermore we have proved the theorem in sec.4 only in even dimensions.
We hope to come back to these problems in a future publication.

{\it Note:}
An independent calculation of the Wodzicki residue in 4 dimensions, leading to
the same result, was recently done by  D.~Kastler \cite{Kastler}.

{\bf Acknowledgments}\\
We would like to thank F.~Scheck, the theory group in Mainz for many helpful
discussions, R.~Coquereaux and the theory group of the C.N.R.S. in Marseille
for a pleasant stay with stimulating discussions.
One of us, M.W., thanks the Graduierten Kolleg, Titel DFG-Kl 681/2-1,
for financial support.

\bye